\documentclass[%
 reprint,
 amsmath,amssymb,
 aps,
]{revtex4-2}

\usepackage{hyperref}
\usepackage{graphicx}
\usepackage{dcolumn}
\usepackage{bm}
\usepackage[utf8]{inputenc}
\usepackage{float}
\usepackage[T1]{fontenc}
\usepackage{color,soul}
\usepackage{mathtools}
\usepackage{epsfig}
\usepackage{color}
\usepackage{url}
\usepackage{siunitx}
\usepackage{enumitem}
\usepackage[english]{babel}
\usepackage{amsmath}
\usepackage{amssymb}
\usepackage{subfig}
\usepackage{bm}
\usepackage{mathptmx}
\usepackage{multirow}
\usepackage{pgf}

\DeclareSIUnit{\belmilliwatt}{Bm}
\DeclareSIUnit{\dBm}{\deci\belmilliwatt}

\begin{document}

\preprint{PRB}

\title[T]{Towards synchronization of serially connected Spin Torque Oscillators based on Magnetic Tunnel Junctions in large field regime}

\author{Piotr Rzeszut$^{a}$}
    \email{piotrva@agh.edu.pl}
    \altaffiliation{equal contribution}
\author{Jakub Mojsiejuk$^{a}$}%
    \email{mojsieju@agh.edu.pl}
    \altaffiliation{equal contribution}
\author{Witold Skowroński$^{a}$}
    \email{skowron@agh.edu.pl}
    \altaffiliation{equal contribution}
\author{Sumito Tsunegi$^{ b}$}
\author{Kay Yakushiji$^{ b}$}
\author{Hitoshi Kubota$^{ b}$}
\author{Shinji Yuasa$^{ b}$}

 \affiliation{a) AGH University of Kraków, Institute of Electronics, al. Mickiewicza 30, 30-059 Kraków, Poland\\
 b) Research Center for Emerging Computing Technologies, National Institute of Advanced Industrial Science and Technology (AIST), Tsukuba, Ibaraki 305-8568, Japan
\\
 }

\date{\today}

\begin{abstract}
Multiple neuromorphic applications require the tuning of two or more devices to a common signal. Various types of neuromorphic computation can be realized using spintronic oscillators, where the DC current induces magnetization precession, which turns into an AC voltage generator. However, in spintronics, synchronization of two oscillators using a DC signal remains a challenging problem, as the oscillations are very sensitive to the device parameter distribution, which inevitably happens during the fabrication process. In this work, we present experimental results on the mechanisms of synchronization of spin-torque oscillators. Devices are based on the magnetic tunnel junction (MTJ) with a perpendicularly magnetized free layer and take advantage of a uniform magnetization precession in the presence of the magnetic field and a DC bias. Oscillations of in-series connection of two MTJs are shown together with the discussion of the possible path towards improving oscillation power and linewidth. To explain the desynchronization of two serially coupled nonidentical oscillators in the experiment, we leverage a simple numerical model. We further show that the in-plane field component and strong field-like torque, along with an adequate arrangement of the oscillators that accounts for magnetic parameters, are key factors for observing desynchronization. Finally, we propose a device in which voltage-controlled anisotropy (VCMA) manipulates the anisotropy of one of the MTJs so that it falls into and out of the synchronized state, which can be further used as a building block of a non-linear computing platform.
\end{abstract}

\keywords{MTJ, Magnetic Tunnel Junction, Synchronization, Neuromorphic computing, Spin torque oscillator, STO}
\maketitle

\section{Introduction}

In recent years, neuromorphic computing concepts have gained rapid traction in multiple disciplines \cite{roy_towards_2019}, with a promising mix of low-power and low-latency hardware embedded algorithms. Exciting work using so-called vortex nanooscillators, such as those by Romera et al. \cite{romera_vowel_2018, romera_binding_2022}, Torrejon et al. \cite{torrejon_neuromorphic_2017} or Tsenugi et al. \cite{tsunegi_scaling_2018} as well as uniform magnetization precession oscillators \cite{Leroux_2022} has also put neuromorphic concepts on the spintronics roadmap \cite{grollier2016spintronic}. In a similar fashion, neuromorphic vowel recognition has been achieved with spin-Hall oscillators based on nanoconstriction arrays \cite{zahedinejad_two-dimensional_2020}. More recently, both experimental and theoretical explorations \cite{houshang_phase-binarized_2022} indicate that spin Hall nano oscillators can be used to construct phase-binarised Ising machines and solve some optimization problems at a potentially lower power consumption than quantum computers.
A general approach involves some type of nanooscillator, where the synchronization mechanism is achieved by a combination of dipole coupling and injection locking \cite{Grollier2020}. 
However, there have been some attempts to achieve coupling based on electric effects rather than magnetic ones, as explored in the work by Taniguchi \cite{taniguchi_synchronization_2020}, which also enabled further tuning. Recently, Sharma et al. \cite{sharma_electrically_2021} have achieved a synchronized oscillation of up to four in-plane magnetized magnetic tunnel junctions (MTJs) connected in parallel or serially. Experimentally, the magnetic isolation of two devices is achieved simply by placing them further apart, so that the dipole interaction between them can be neglected (the dipole interaction decreases with distance as $d^3$). This usually requires an inter-device spacing of the order of $\SI{}{\micro m}$.

In this work, we present a numerical and experimental study of in-series connected spin torque oscillators (STOs) based on MTJ with mixed magnetic anisotropies. To maximize the spin transfer torque (STT) acting on the free layer, an MTJ is used with an in-plane magnetized reference layer and a perpendicularly magnetized free layer \cite{Houssameddine2007}. At optimal magnetic field orientation and DC bias conditions, this MTJ operates as an STO at radio frequencies with an average Q factor of 400. Next, synchronization of the STO to both external reference signal and the second, MTJ-based STO is shown. 
Oscillations of synchronized and desynchronized in-series systems are analyzed in detail, using numerical simulations, exhibiting asymmetric influence of the magnetic anisotropy variations on the oscillations conditions. Finally, we propose a device, where the synchronization state is determined based on voltage controlled magnetic anisotropy (VCMA) \cite{maruyama_large_2009}.

\section{Experimental methods}
The MTJ multilayer structure used for STO is as follows: Si /SiO$_\mathrm{2}$ / buffer / (4) Ru / (6) IrMn / (2.5) CoFe / (0.8) Ru / (2.5) CoFeB / (1) MgO / (1.8) CoFeB / (1) MgO / (3) Ru / (5) Ta / (2) Ru / (3) Pt (thickness in \si{\nano\metre}). 
The bottom CoFeB reference layer is in-plane magnetized and is coupled to the CoFe layer, which is pinned to the IrMn antiferromagnet. The top CoFeB layer is characterized by a perpendicular magnetic anisotropy\cite{ikeda_perpendicular-anisotropy_2010}, due to the significant contribution of the double MgO interface anisotropy \cite{skowronski_2018_influence}. The thickness of the MgO tunnel barrier corresponds to the resistance-area product of \SI{3.6}{\ohm\micro\metre\squared}. The buffer layer consists of Ta/Cu/Ta, which has small surface roughness.

\begin{figure}[ht]
    \centering
    \includegraphics[width=\columnwidth]{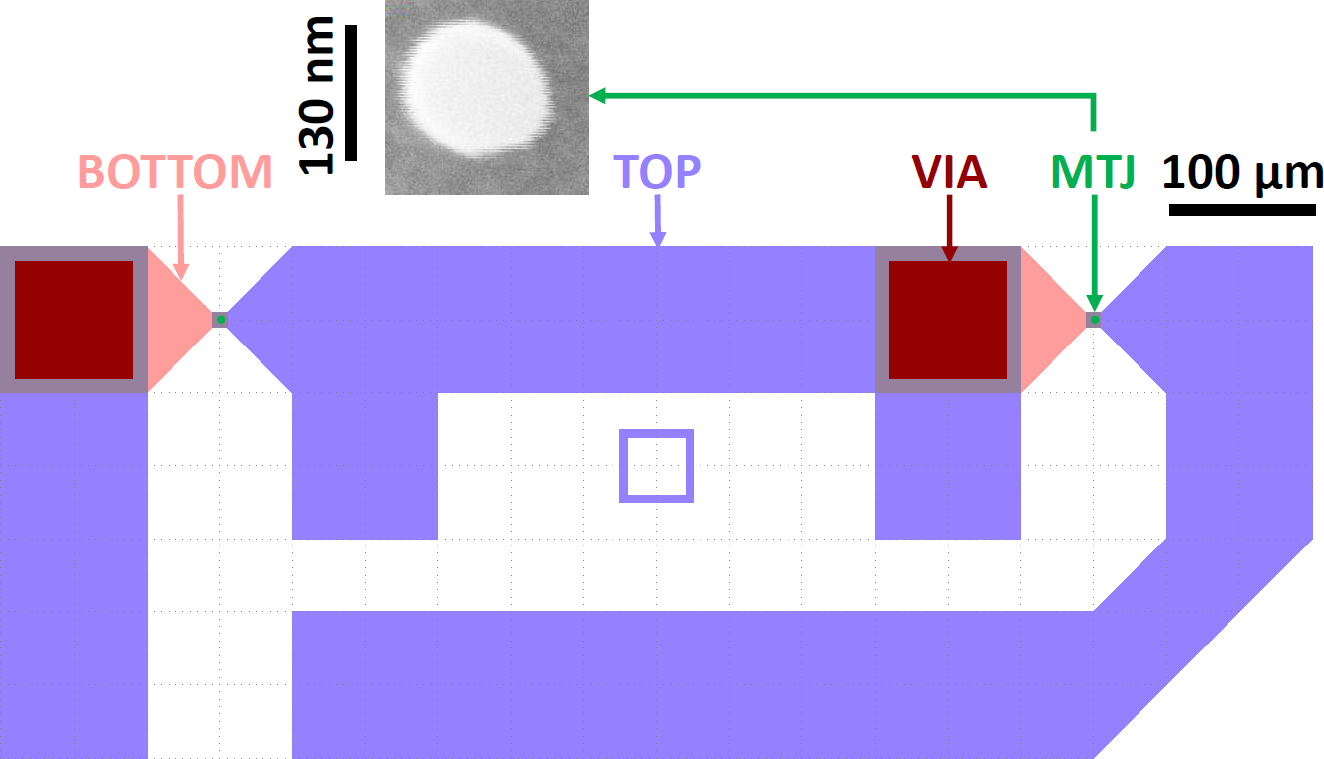}
    \caption{Lithography mask of a single device consisting of two serially (head-to-tail) connected MTJs. Pink color denotes a bottom electrode, red the via, violet represents the top electrode. MTJs are not visible at this scale and are fabricated on thin intersections of top and bottom electrode.}
    \label{fig:sample-mask}
\end{figure}

Mixed match three-step lithography is performed using electron and optical lithography, including ion beam etching and the lift-off process to fabricate serially connected MTJs with the necessary measurement contacts (Fig.\ref{fig:sample-mask}). The nominal dimensions of all MTJs are \SI{130}{\nano\metre} in diameter, which are encapsulated in $\mathrm{Al}_\mathrm{2}\mathrm{O}_\mathrm{3}$. The resistance of a single MTJ is around \SI{67}{\ohm}. A detailed magnetic characterization and fabrication process is described in our earlier work \cite{skowronski_microwave_2019}. 

Both static and RF characterization of the fabricated devices were performed in a dedicated probe station equipped with the electromagnet and rotating arm with broadband cabling, enabling measurement of the STO signal in the presence of a static magnetic field at an arbitrary angle and DC bias supplied from a precise source meters. Details of the measurement setup are given in Appendix \ref{app:measurement}. We used several pairs of MTJs for the current study, which show similar properties.

\section{Results and discussion}
 \begin{figure*}
    \centering
    \includegraphics[width=\textwidth]{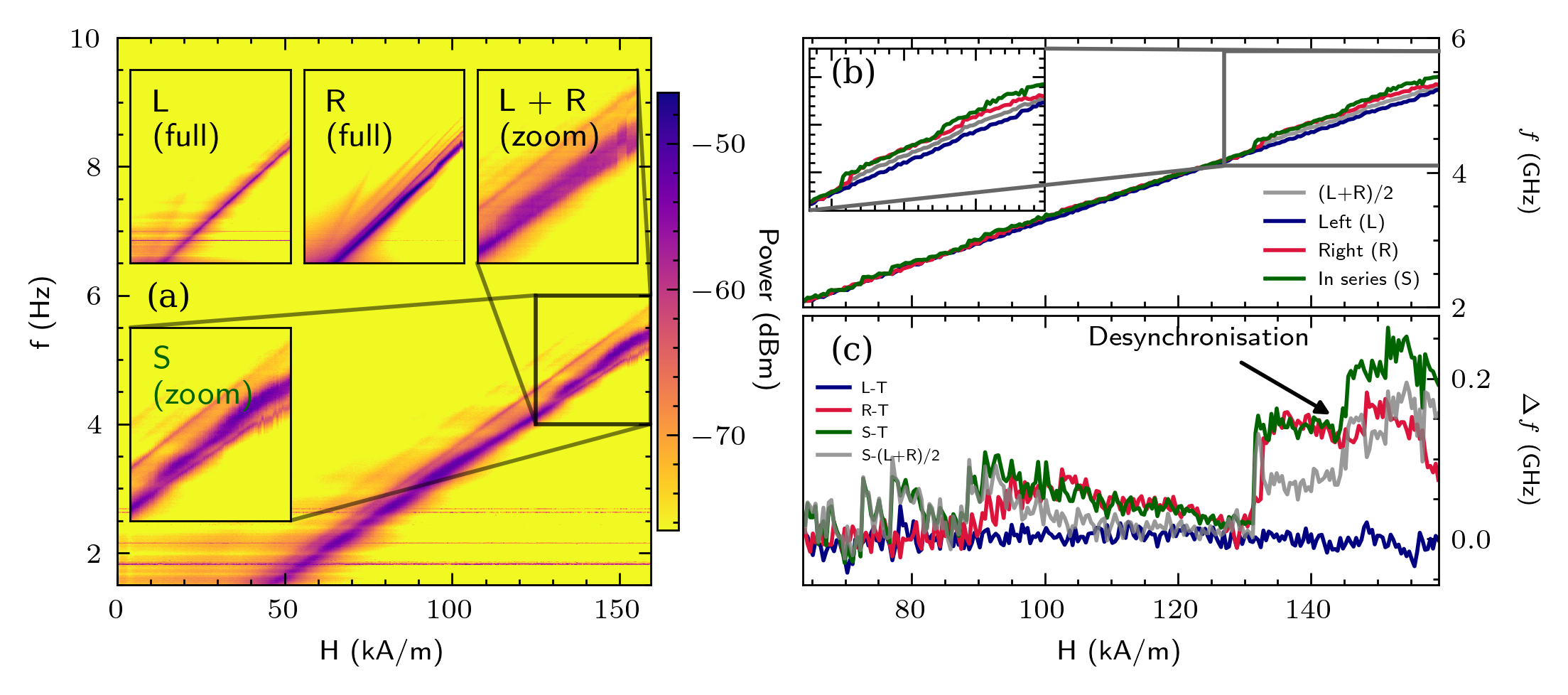}
    \caption{Experimental electrical frequency synchronization and desynchronization of two serially connected MTJs. (a) In-series signal with insets illustrating separate left (L) and right (R) MTJ in the entire field region (0, 155) $\si{kA/m}$, and a mathematical average of left and right (L + R) compared with a zoom of the in-series (S), both zoomed in the range (125, 155) $\si{kA/m}$. In the zoomed region, we can observe a line with a small FWHM separating from a wide mode; a similar picture does not emerge simply by adding the left and right signals together. (b) Main oscillating frequency of the L, R and in series systems, with a zoom of the desynchronization region.
    (c) the synchronization and desynchronization ranges of L and R and in-series (S) signal respective to the trend line (T) fitted to the L signal. L signal should, in theory, be least affected by the in-series electrical connection, and therefore is used as a reference. The arrow marks the spot where the R signal diverges from L, and S diverges both from L and R. As shown, up until about 145 kA/m, the S signal follows R closely, then settles to oscillate with its own distinct frequency. In all panels we also include a mathematical average of (L) and (R) signals, (L+R)/2, for reference. In (c) that signal is subtracted from (S) for comparison.}
    \label{fig:series}
\end{figure*}
\subsection{Synchronization of two serially connected STO}
A serial connection of two MTJs such as the one shown in Fig.\ref{fig:sample-mask} implies that the first MTJ in series (let us call it \textit{L}) will act as a source of an injected current signal to the second MTJ (\textit{R}). Therefore, it is useful to first consider the ability of the fabricated MTJ to synchronize with an external generator signal. In Appendix \ref{app:external-synchronization} we discuss the details of such a synchronization based on the measurements of one of the MTJs in the series. As a result, we show the ability to lock the STO signal to twice the oscillation frequency fed by the generator.

The measurements of two serially connected STOs were performed in the following way: we collected auto-oscillation data for each element separately (indicated as left (L) and right (R) in Fig.\ref{fig:series}), and then measured them when connected in series with a tail-to-head type connection. In all cases, the current flowing through the elements was kept at the same value: $I_{DC} = \SI{-2.90}{\milli\ampere}$. For each situation, we calculated and compared separate elements with their serial connection, as presented in Fig.\ref{fig:series}a-c). As a result, both MTJs oscillate under similar conditions, especially at magnetic fields below 1.6 kA/m; therefore, to determine the synchronization conditions, one needs to analyze the oscillation linewidth (quantified as a full width at half maximum - FWHM). In the simulation part, we begin by introducing the model and fitting parameters for a single MTJ. Then, we used an external RF source to determine the locking range of the MTJ oscillator, which we later modeled in the simulated oscillations of coupled STOs.

We move on to a detailed discussion of Fig.\ref{fig:series}, specifically the separation of the oscillating frequencies at about 145 kA/m. The first observation is that the oscillation pattern shown in Fig.\ref{fig:series}a) is more complex for a serial connection than for a single element  synchronizing with the $2f$ signal (presented  for comparison in Appendix \ref{app:external-synchronization}). In particular, the inset panels of the (\textit{R}) oscillation, as well as the zoomed-in inset of (\textit{S}) in Fig.\ref{fig:series}a) show a set of weaker oscillating frequencies in proximity of the main one, which can be contrasted with a relatively clean oscillating relation of the \textit{L} signal, also in Fig.\ref{fig:series}a). This particular feature is later replicated with the simulated spectra, cf. Fig.\ref{fig:wide-panel-sim}b).
Fig.\ref{fig:series}c) reveals more information on the differences between the in series oscillation (\textit{S}) and the individual main oscillating modes of the (\textit{L}) or (\textit{R}) elements measured when they were not connected serially. Each line represents a difference between a main oscillating frequency (taken from Fig.\ref{fig:series}b)) and a trend line (\textit{T}). The trend line was fitted to the signal of (\textit{L}) as seen in Fig.\ref{fig:series}b), because that spectrum is principally the least affected by the in series electrical connection.  The trend line is defined as a straight line of the form$f = k H + H_0$, where $k$ is the slope, and $H_0$ is the intercept. The trend line was introduced for visual presentation purposes, and helps to clean out micro irregularities from the main resonance line of $L$. Starting from about 95 \si{kA/m} we see that the line (\textit{R}-\textit{T}) follows the line (\textit{S}-\textit{T}), which means that the in-series oscillator aligns itself with one of the individual modes (\textit{R}) measured outside the in-series connection, because (\textit{L}-\textit{T}) lingers near 0 during the same field range This situation continues, even in the region, when at approximately 132 \si{kA/m} the (\textit{R}) and (\textit{L}) signals begin to drift further apart from each other. In between, we observe a brief moment of non-forced synchronization in the region 125\si{kA/m}, where all signals naturally oscillate with the same frequency.
Finally, beyond the 145 \si{kA/m} mark, the line (\textit{S}-\textit{T}) does not align with neither (\textit{R}-\textit{T}) nor (\textit{L}-\textit{T}). Such a pseudo-desynchronization from the (\textit{R}) signal means that the in-series system begins to oscillate at its own individual frequency, which is different from the one enforced by (\textit{R}) or \textit{L}. At the same time, in Fig.\ref{fig:series}a) we observe a separation of a strong main oscillating frequency line, clearly visible in the zoomed inset (\textit{S}) for the range starting from 125 \si{kA/m}. When we average the two individual signals from (\textit{L}) and (\textit{R}) (see inset titled (\textit{L} + \textit{R}) in Fig.\ref
{fig:series}b)), this mode is not clearly distinguishable as a contribution from either of the MTJs. Similarly, subtracting a mathematical average (\textit{L}+\textit{R})/2 from (\textit{S}) yields a distinctively different behavior when compared with (\textit{S}-\textit{T}), as shown in Fig.\ref
{fig:series}c). To gain a deeper understanding of the mechanisms of synchronization, we performed numerical simulations of serially connected MTJs.

\subsection{Numerical simulations}
In our numerical simulations, we use the Landau-Lifschitz-Gilbert-Slonczewski (LLGS) macrospin equation \cite{gilbert_classics_2004,ralph_spin_2008,slonczewski_current-driven_1996,slonczewski_currents_2002}:

\begin{multline}
\frac{1+ \alpha_\textrm{G}^2}{\gamma_0}\frac{\textrm{d}\mathbf{m}}{\textrm{dt}} = -\mathbf{m} \times \mathbf{H}_{\mathrm{eff}} - \alpha_\textrm{G}\mathbf{m}\times\mathbf{m}\times\mathbf{H}_{\mathrm{eff}}  \\
+ a_j \epsilon\beta \mathbf{m} \times \mathbf{p}
-a_j\epsilon (\mathbf{m}\times\mathbf{m}\times\mathbf{p})
\label{eq:numerical-ll-stt}
\end{multline}

where $\mathbf{m} = \frac{\mathbf{M}}{M_s}$ is a normalized magnetization vector, with $M_\textrm{s}$ as the magnetization saturation, $\alpha_\textrm{G}$ as the Gilbert damping parameter, $\gamma_{0}$ is the gyromagnetic factor and $\mathbf{H}_\mathrm
{eff}$ is the effective field vector. 
For our simulations, we include the spin-torque terms with $\beta$ as the scaling factor for the field-like torque, and $\mathbf{p}$ as the vector of the normalized reference layer:
\begin{equation}
    a_j = \frac{\hbar j}{e \mu_0 M_\textrm{s} t_\textrm{FM}}
\end{equation}
where $j$ is the input current density through the reference layer, $t_\textrm{FM}$ is the thickness of the ferromagnetic layer (FM), $e$ is the electron charge, and $\epsilon$ is defined as:
\begin{equation}
    \epsilon = \frac{P \lambda^2}{\lambda^2 + 1 + (\lambda^2 - 1)(\mathbf{m}\cdot \mathbf{p})}
\end{equation}
where $\lambda$ is the angular parameter of the STT, and $P$ is the polarisation efficiency.

In our experiments, we include the following main contributions to the effective field: the demagnetization field, the anisotropy field, and the external magnetic field. Each MTJ is made up of two FM layers, where we assume that the bottom layer is a fixed reference layer for the top layer, with the polarization vector aligned along the $x$ axis $p_x$.

We fit the experimental data using the simulation package \textsc{cmtj} \cite{mojsiejuk_cmtj_2023} and determine the parameters of one of the MTJs in series as $\mu_0M_\mathrm{s} = 0.75$ T and $K_\mathrm{u} = 0.23 \ \mathrm{MJ/m^3}$, with the perpendicular anisotropy axis. For consecutive simulations, we use the constant current density of $j_\mathrm{L} = 22 \ \mathrm{MA/cm^2}$, $\lambda^2 = 0.28, \alpha_\mathrm{G} = 0.01$, which are typical parameter values for STOs \cite{Costa2017}.

\subsection{Model of synchronization}


The synchronization behavior of electrically coupled MTJs is enforced after Taniguchi et al. \cite{taniguchi_synchronization_2020}. We assume that two oscillators are sufficiently far apart so that there is no dipole interaction between them.
There are two available configurations for connecting two MTJs, serial and parallel, and in this work we only consider the former. In each, the current depends on the free magnetizations $\mathbf{m}$ and the reference layers $\mathbf{p}$ of the junctions $i_{th}$ and $i+1_{th}$:
\begin{align}
    \label{eq:coupling}
    I(t) = I_0(t) + \chi I_0(t)(\mathbf{m_i} \cdot \mathbf{p}_i + \mathbf{m}_{i+1} \cdot \mathbf{p}_{i+1}) \quad \textrm{(series)}\\ 
    I(t) = I_0(t) + \chi I_0(t)(\mathbf{m_i} \cdot \mathbf{p}_i - \mathbf{m}_{i+1} \cdot \mathbf{p}_{i+1}) \quad \textrm{(parallel)}
\end{align}

where $I_0(t)$ is the value of the current fed into the first MTJ in the connection and $\chi$ is the unitless coupling strength. For convenience, the effective coupling is defined as $\chi' = \chi (m_\mathrm{x, L} + m_\mathrm{x, R})$ as both reference layers are defined in the simulation to be along the $x$ axis.
    
\begin{figure*}
    \centering
    \includegraphics[width=\textwidth]{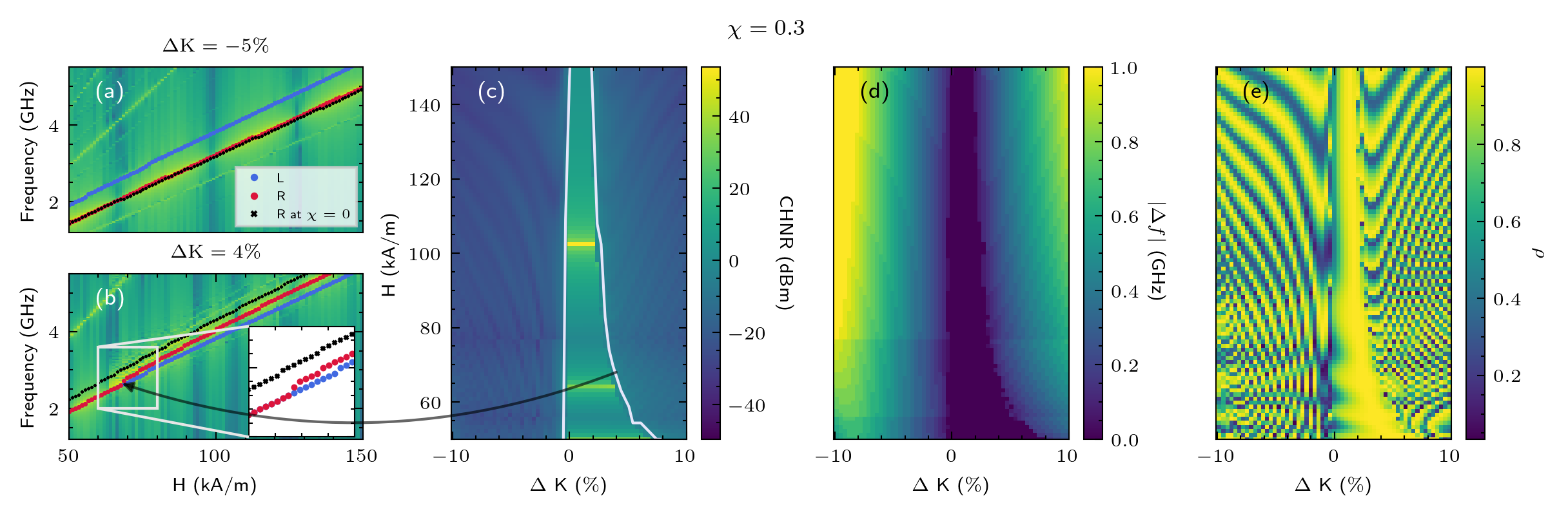}
    \caption{Simulated desynchronization of serially coupled MTJs in function of the anisotropy change $\Delta \mathrm{K}$ of the (R) MTJ in series. 
    Panels (a) and (b) show two sample spectra for R junction taken at $\Delta \mathrm{K} = -5 \%$ (a) and $\Delta \mathrm{K} = 4\%$ (b). The arrow leads from the desynchronization boundary at $\approx 68 ~ \mathrm{kA/m}$ and $\Delta \mathrm{K} = 4\%$ in (c) to a precise place of desynchronization in the spectrum of (b), with the desynchronization moment placed in zoom in (b). (a-b) The blue marker follows the line of the main oscillating frequency of L, and the red marker tracks the one for R. The black dots show the main oscillating frequency line of R when the two MTJs are decoupled, i.e. $\chi = 0$. Otherwise, simulations performed assuming $\chi = 0.3$. We used $\beta = 0.5$.
    (d) Frequency difference in (GHz) between the L and R junctions. For $\Delta \mathrm{K} > 0$, i.e. $K_\mathrm{R} > K_\mathrm{L}$ there is a desynchronization boundary where the L and R junctions start to fall out of synchronization, i.e. leave the dark blue region (white line in (c) denotes that desynchronization boundary $H_\mathrm{thres}$). (c) illustrates that the dark-blue region from (d) is characterized by a large CHNR, supporting the evidence for strong synchronization. The asymmetry between the frequency de-synchronization fields for $\Delta K > 0$ and $\Delta K < 0$ arise from the particular arrangement of the (L) and (R) MTJs in series, and is a direct consequence of a smaller frequency changes due to current change. High order parameter values (e) confirm a good degree of synchronization within the desynchronization boundary.}
    \label{fig:wide-panel-sim}
\end{figure*}

We now move to a qualitative demonstration of a desynchronization in a numerical setting. Contrary to previous works \cite{taniguchi_mutual_2018,arun_influence_2020} considering electrically coupled MTJs, we assume that there exists a nontrivial parameter dispersion between two MTJs, that is, they are not identical. Having two junctions connected in series, the magnetic field angle is set at $\theta = 25^\circ$, the same as in the experiment. We scan with the anisotropy value of the second MTJ in the connection, a variation of the anisotropy with respect to the first MTJ in the connection is $\Delta \mathrm{K}$, such that $K_\mathrm{u, R} = K_\mathrm{u, L} + \Delta \mathrm{K}$. For every field magnitude and $\Delta \mathrm{K}$ value, we note the frequency difference of the primary modes between the first and second MTJ, which results in Fig.\ref{fig:wide-panel-sim}b).

\begin{figure*}
    \centering
    \includegraphics[width=\textwidth]{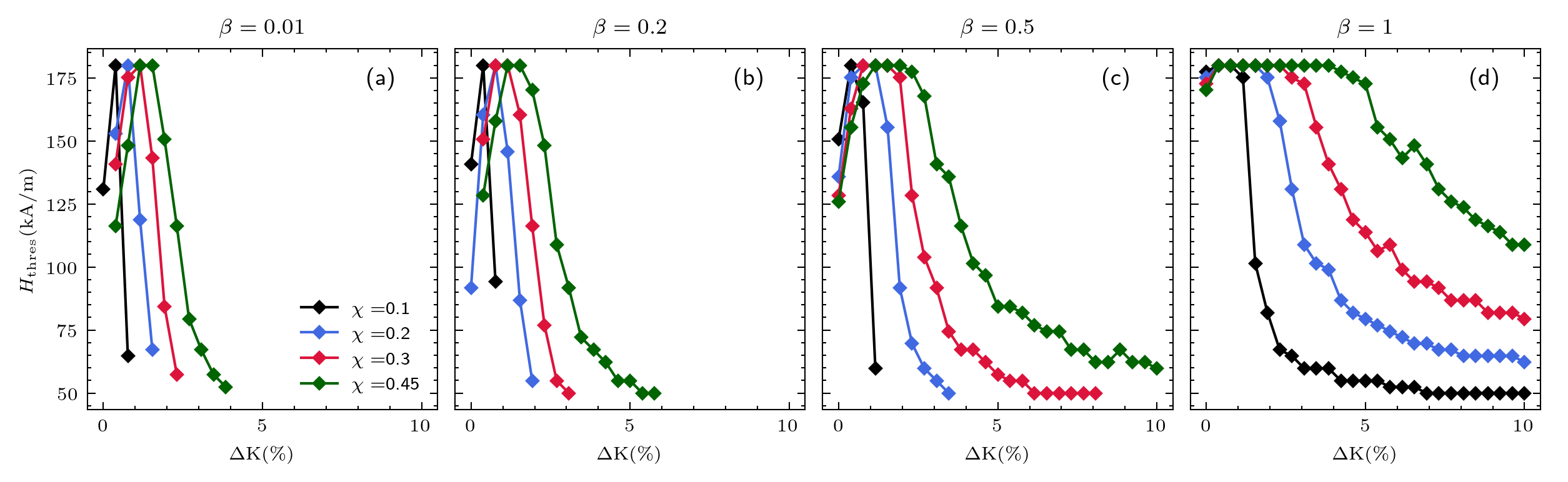}
    \caption{The effect of field-like torque scaling $\beta$ and the magnitude of the electric coupling $\chi$ on the desynchronization threshold $\mathrm{H}_\mathrm{thres}$ in function of the anisotropy change $\Delta \mathrm{K}$ of the (R) MTJ in series. Case (a) corresponds to $\beta = \alpha_\mathrm{G}$. Increasing $\beta$ positively correlates with the increased synchronization range across magnetic parameter dispersion, having a similar effect to increasing the coupling constant $\chi$. External field range is capped to 175 kA/m. Plotted points denote only states that were synchronized at some field, a lack of a marker means that the state either did not start in a synchronized state at lower field values or it has already desynchronized.}
    \label{fig:beta-changes}
\end{figure*}

The distribution of the synchronized states shifts off-center towards $\Delta \mathrm{K} > 0$ and then decreases with a slope dependent on the $\chi$, the coupling strength. Vertically, at some $H_\mathrm{thres}$, marked with a white line in Fig.\ref{fig:wide-panel-sim}c), there is a transition from the synchronized state to the desynchronized state. To quantify the strength of the synchronization, we define cross-Harmonic-to-Noise (CHNR) ratio as:
\begin{equation}
    \mathrm{CHNR} = \frac{\sum_{\forall n > 0, n \in \mathbf{Z}} \mathnormal{A}^\mathrm{R}_{nf_\mathrm{L}}}{\sum_{f' \neq kf_\mathrm{L} \forall k >1, k \in \mathbf{Z} } \mathnormal{A}^\mathrm{R}_{f'}}
\end{equation}
where $f_\mathrm{L}$ is the primary oscillating frequency of the first MTJ in series. $\mathnormal{A}^\mathrm{R}_{f_i}$ denotes the values taken from the frequency spectrum of the second MTJ in series, that is, the amplitude at a frequency $f_i$. Fig.\ref{fig:wide-panel-sim}c) shows the CHNR where the transition to the desynchronized state is connected with an abrupt nonharmonic oscillation of the second MTJ, which results in frequency splitting similar to the signal observed experimentally in Fig.\ref{fig:series}a), panel (\textit{R}) -- numerical equivalent presented in Fig.\ref{fig:wide-panel-sim}b). As in the experiment, the main oscillating mode follows neither (\textit{L}) nor (\textit{R}) modes exactly.
In addition to CHNR, we consider order parameter as an additional measure of uniformity of synchronization \cite{acebron_kuramoto_2005,kanao_reservoir_2019}:
\begin{equation}
    \rho = \frac{1}{2}\left|\sum_i^2 \frac{c_i}{|c_i|}\right|
\end{equation}
where $c_i = m_x^i - i m_y^i$ with $m_x$ and $m_y$ as the $x$ and $y$ components of the magnetization. The closer $\rho$ approaches 1, the better the order of synchronization. The results are shown in Fig.\ref{fig:wide-panel-sim}e), where for the synchronized region we clearly observe swaths of large $\rho$ values.

The mechanism of desynchronization we propose is based on the fact that for the MTJs in question, the increase in current density supplied to a device causes a decrease in frequency of oscillation (and vice versa). We further note that the input current density depends solely on (\ref{eq:coupling}), and therefore on $m_{x, L}$ and $m_{x, R}$ components. Oscillation frequency of (\textit{L}) and (\textit{R}) is denoted as as $f_\mathrm{R}$ and $f_\mathrm{L}$ respectively. In the following, we discuss two specific cases: 

(1) When $\Delta \mathrm{K} > 0$ then the oscillating frequency of (\textit{R}) is larger than that of (\textit{L}) for the same applied field value, and $m_{x, L}, m_{x, R}$ are both small in the low $H_z$ regime. Consequently, $\chi'$ is relatively large, allowing for stronger injection locking. As $H_z$ increases, so does $m_{z, L}, m_{z, R}$ at the cost of $m_{x, L}, m_{x, R}$, which in turn leads to a drop in $\chi'$, and thus $j_\mathrm{R}(t)$, the current density supplied to the \textit{R}. 
The drop in $j_\mathrm{R}$ causes $f_\mathrm{R}$ to increase, driving even further from $f_\mathrm{L}$, and this continues until $f_\mathrm{R}$ falls outside the locking range for a given $K_\mathrm{u, R}$. This process is accelerated as $\Delta \mathrm{K}$ grows larger, because $\frac{\mathrm{d}f}{\mathrm{d}j}$ increases with $\Delta \mathrm{K}$, as well as the nominal difference between $f_\mathrm{R}, f_\mathrm{L}$ in a decoupled case -- those factors account for the observed slope of $H_\mathrm{thres}$ for $\Delta \mathrm{K} > 0$.

(2) In the case of $\Delta \mathrm{K} < 0$, $m_{x, R}$ starts out larger due to lower $K_\mathrm{u}$, and in small fields it means an increase in $\chi'$. Because in the decoupled case we have that $f_\mathrm{L}$ > $f_\mathrm{R}$, this increase in $\chi'$ causes $f_\mathrm{R}$ to move away from $f_\mathrm{L}$, in an opposite direction to the one discussed above, when $\Delta \mathrm{K} > 0$, due to an increase in the current density flowing to the second MTJ in series. The smaller $\frac{\mathrm{d}f}{\mathrm{d}j}$ for $\Delta \mathrm{K} < 0$ means that the possibility of current regulation via coupling is also much smaller. This further suggests why, given the initial conditions, the states for $\Delta \mathrm{K} > 0$ begin to synchronize at low $H_z$. 

\begin{figure*}
    \centering    \includegraphics[width=\textwidth]{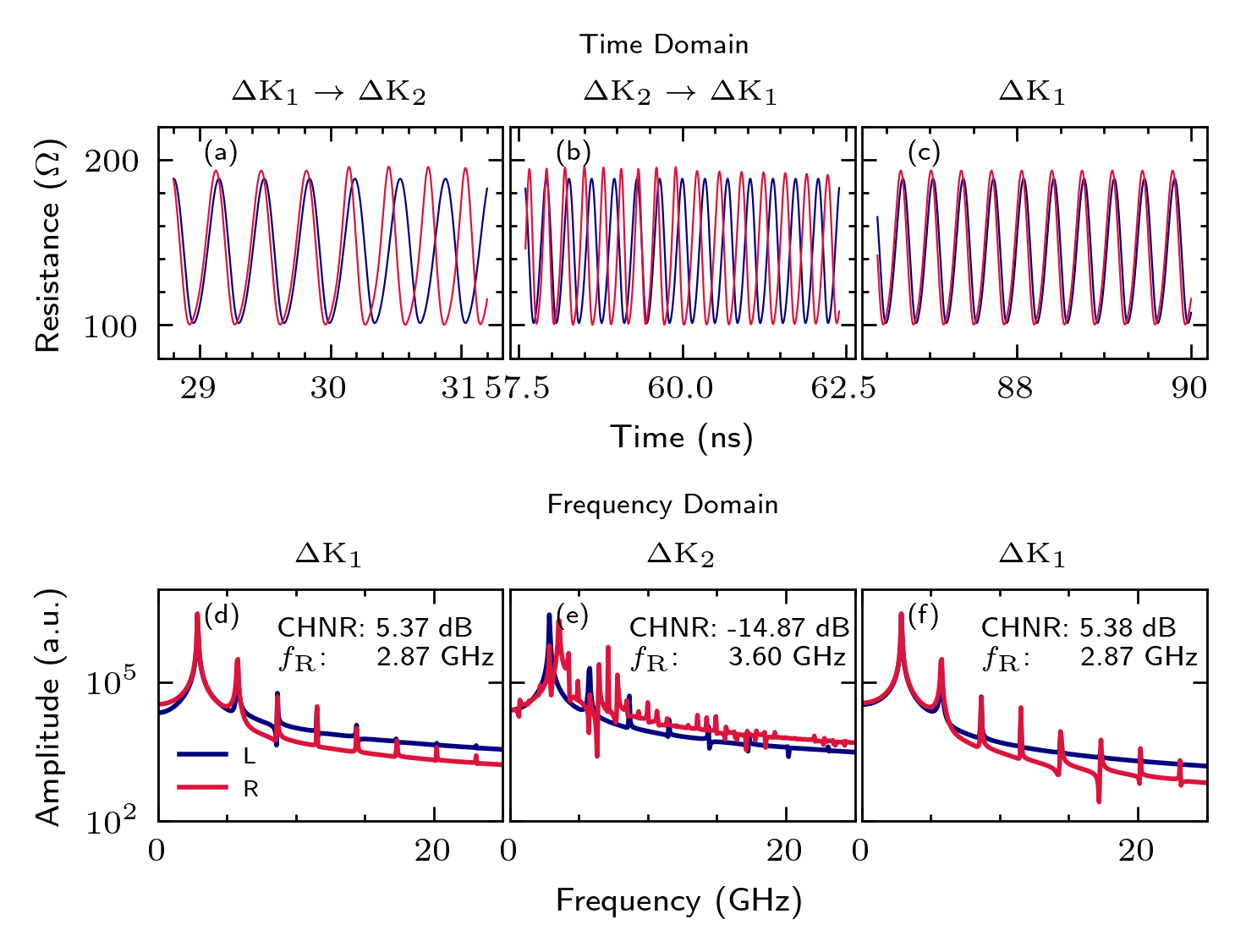}
    \caption{A simulated VCMA-controlled synchronization of and in-series MTJ system. The VCMA controller switches from $\Delta \mathrm{K}_1$ to $\Delta \mathrm{K}_2$ first, then back to $\Delta \mathrm{K}_1$. (a) shows the magnetoresistance during transition at 30 ns from $\Delta \mathrm{K}_1$ to $\Delta \mathrm{K}_2$, (b) at 60 ns from $\Delta \mathrm{K}_2$ back to $\Delta \mathrm{K}_1$ and (c) gives last couple nanoseconds from a synchronized state in $\Delta \mathrm{K}_1$ after the final switch. Frequency spectra corresponding to states $\Delta \mathrm{K}_1$, $\Delta \mathrm{K}_2$ and $\Delta \mathrm{K}_1$ (after the transition from $\Delta \mathrm{K}_2$) are shown in (d-f) respectively. Simulated for $\chi = 0.15$, $\Delta \mathrm{K}_1 = 4\%$, $\Delta \mathrm{K}_2 = 20\%$, $H_\mathrm{ext} = 80 \mathrm{kA/m}$.}
    \label{fig:vcma-example}
\end{figure*}

A similar analysis can be performed when $M_\mathrm{s, R}$, or both $M_\mathrm{s, R}$ and $K_\mathrm{u, R}$ are varied. In the former case, the effect of $\Delta M_\mathrm{s}$ is a mirror reflection along the $y$ axis of the plot presented in Fig.\ref{fig:wide-panel-sim}d), as increasing $M_\mathrm{s, R}$ has the opposite effect of increasing the perpendicular anisotropy $K_\mathrm{u, R}$. Likewise, for the latter combination, one can balance the two magnetic parameters following the equilibrium line $K_\mathrm{u, R} = M_\mathrm{s, R}^2/{2\mu_0}$.

We want to briefly touch on the importance of choosing $\beta$ for desynchronisation fields. As pointed out by Arun et al.\cite{arun_influence_2020}, the field-like torque term plays a significant role in increasing the likelihood of obtaining the synchronized state. In the context of Fig.\ref{fig:wide-panel-sim}, one could expect the widening of the synchronized region, alternatively, a movement of the desynchronization $H_\mathrm{thres}$ towards larger values of $\Delta K$. In fact, we observe that pattern, where increasing $\beta$ in (\ref{eq:numerical-ll-stt}) leads to higher values of the desynchronization threshold field, as shown in Fig.\ref{fig:beta-changes}. As a further extension of the threshold desynchronization field in Fig.\ref{fig:wide-panel-sim}c), Fig.\ref{fig:beta-changes} demonstrates the widening of the synchronization region as a function of increasing $\chi$. We also note the presence of an off-zero peak visible for lower $\beta$ values, which suggests that the longer field-locking ranges for a non-negative $\Delta \mathrm{K}$ may not always occur exclusively when the two MTJs have identical parameters.

The direct consequence of the existence of the threshold field is that it is theoretically possible to construct a device in which VCMA \cite{nozaki_recent_2019, Gonzalez2022} manipulates the second (\textit{R}) MTJ anisotropy such that it falls into and out of the synchronized state. As before, we simulate two devices (\textit{L} and \textit{R}) connected in series. The oscillation of \textit{L} is induced by either the STT mechanism or by the spin orbit torque (SOT) if the current is fed through a heavy metal layer placed under the ferromagnetic one. The electric coupling modifies the current entering the \textit{R} device which oscillates with its own pattern, also by STT or SOT. In the following steps, we alter the perpendicular anisotropy constant of the \textit{R} device via the VCMA to cause a desired sequence of synchronized or desynchronized states. The (\textit{R}) MTJ starts with $\Delta \mathrm{K}_1 = 4\%$, which after Fig.\ref{fig:injection-spectra} guarantees synchronization at $80 \mathrm{kA/m}$. Then, using VCMA we increase the anisotropy constant of \textit{R} to $\Delta \mathrm{K}_2 = 20\%$, and MTJ experiences almost immediate desynchronization, as shown in Fig.\ref{fig:vcma-example}(a), where a mark of 30 ns designates the $K$ change. This is further supported by the frequency spectrum in Fig.\ref{fig:vcma-example}(e). At 60 ns, we bring the anisotropy of \textit{R} down to the $\Delta \mathrm{K}_1$ state, Fig.\ref{fig:vcma-example}b), which after some time achieves synchronization again Figs.\ref{fig:vcma-example}(c) and (f). Such a device may prove useful in efficient computing \cite{houshang_phase-binarized_2022}: a synchronized signal is easily detectable electrically due to an increase in amplitude, and a relatively small voltage is required to tip the pair from the synchronized state.

\section{Conclusions}
Achieving electric synchronization with spin-torque oscillators requires a great level of precision and tedious tuning of the junction parameters. Apart from the difficulty in orchestrating the correct setup, specific synchronization ranges depend largely on the relative parameter dispersion. On the basis of the experimental approach, we were able to present the oscillations in individual STOs and in a serially connected pair of MTJs. We note that the desynchronization, i.e. the relative drift of the individual modes from the common mode, can be both measured experimentally and modeled with a relatively simple numerical model. We further show that the ordering of the junctions in the connection is of significance to the ease of achieving synchronization and the difficulty in achieving the synchronization is non-linearly dependent on the difference between the magnetic parameters of the two junctions constituting the connection. This is interesting specifically in the context of $\Delta \mathrm{K}$, which can be manipulated by the VCMA -- one can find an operating point where, depending on the controlling voltage, serially connected MTJs are moved into and out of the synchronized state.
Further work on achieving reliable locking of two MTJs in the same mode may focus on connecting more devices together, similar to SHNO chains \cite{Litvinenko_2023}, creating less selective circuits with broader bandwidths. Developing such a method can lead to promising opportunities to create arrays of tunable oscillators \cite{Jenkins_2021} capable of performing simple computing tasks. Indeed, our simulations show that by using VCMA to asymmetrically tune the magnetic properties of each STO, one can control the synchronization state of the oscillator chains.

\section{Acknowledgements}

We acknowledge the discussion with dr. Sławomir Ziętek and dr. Jakub Chęciński. The research project is partially supported by the \textit{Excellence initiative – research university} program of the AGH University of Krakow and by the Polish Ministry of Education and Science under subvention funds for the Institute of Electronics of AGH University of Krakow. 
W.S. acknowledges project 2021/40/Q/ST5/00209 from the National Science Center, Poland.

\section{Author contributions}
 S.T., K.Y., H.K. and S.Y. fabricated the sample and provided major suggestions on the manuscript. P.R. and W.S performed the lithography and conducted the measurements. J.M. performed the simulations and formal analysis. P.R., J.M. and W.S. co-wrote the article. All authors contributed to the review and verification of the manuscript.
\newpage
\bibliography{zotero_ref_PR}

\begin{appendix}
\clearpage
\newpage
\section{Measurement setup}
\label{app:measurement}
\begin{figure}[ht]
    \centering
    \includegraphics[width=\columnwidth]{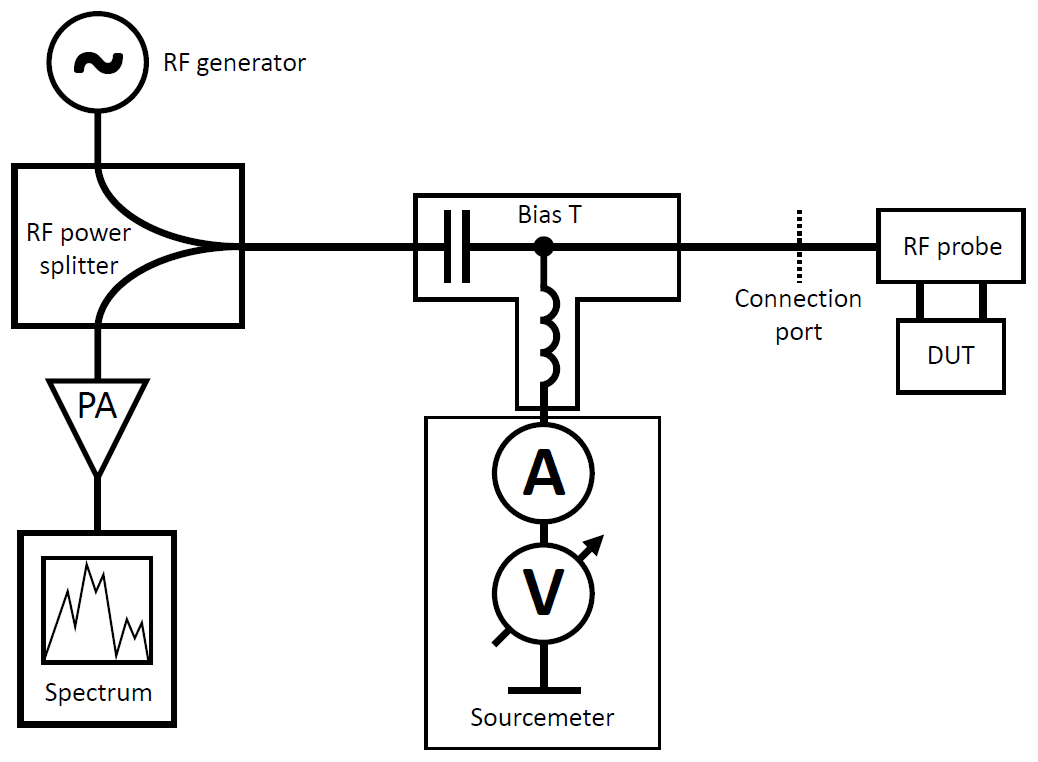}
    \caption{Measurement setup used for injection locking experiment, utilizing directional coupler, RF generator, spectrum analyzer, bias T and sourcemeter.}
    \label{fig:measurement-setup}
\end{figure}

Schematic of the measurement setup used in the experiment is presented in Fig.\ref{fig:measurement-setup}. The setup consists of a radio frequency generator (RF) (Agilent E8257D) connected to the first symmetric port of the RP power splitter (Mini-Circuits ZN2PD2-14W-S+). The second symmetric port of the splitter is connected to the RF power amplifier (Mini-Circuits ZVA-213-S +) powered by a 12V DC power supply. The amplified signal is fed into an RF spectrum analyzer (Agilent N9030A). The combined (main) port of the power splitter is connected to an RF port of a bias tee (Mini-Circuits ZX86-12G-S+). A DC port of the bias tee is connected to the sourcemeter (Keysight B2912A). RF port of the bias tee is connected to an RF probe (Picoprobe) that allows for the connection of the sample. When the RF generator was not used, it was replaced with the \SI{50}{\ohm} RF terminator.

The measurement system was completed with a magnetic field source consisting of an electromagnet (GMW model 3472-70) supplied with a current source (Kepco power supply BOP 50-20MG) and a gaussmeter (Lakeshore DSP-475) for magnetic field readout during electromagnet calibration. To change the orientation of the magnetic field vector relative to the sample plane, a dedicated rotating probe station (manufactured by Measline Ltd.) driven by a specially designed linear stepper motor driver system is used, described in \cite{skowronski_angular_2021}.

\section{Synchronization to an external signal}
\label{app:external-synchronization}
\subsection{Synchronization of a single element to an external $2f$ signal}

Our study of synchronization begins with STO locking oscillations to the external RF signal through the injection locking mechanism, following the procedure described in \cite{quinsat_injection_2011}. To determine how it affects the signal parameters, we devised a setup (Fig.\ref{fig:measurement-setup}) consisting of MTJ connected via a bias-T to a sourcemeter (which provides constant current bias) and to the RF setup: directional coupler, spectrum analyzer, and RF signal generator. As it is generally difficult to obtain ideal impedance matching between the MTJ and probe connection, as the impedance of the MTJ itself is usually different from \SI{50}{\ohm}. This makes observation of direct synchronization to the $1f$ signal difficult due to the significant reflection of the signal from the STO. Therefore, synchronization with the $2f$ signal is used to investigate synchronization of the MTJ with an external excitation, and the full width at half maximum (FWHM) parameter is used as a synchronization indicator. 

\begin{figure*}[ht]
    \centering
    \includegraphics[width=\textwidth]{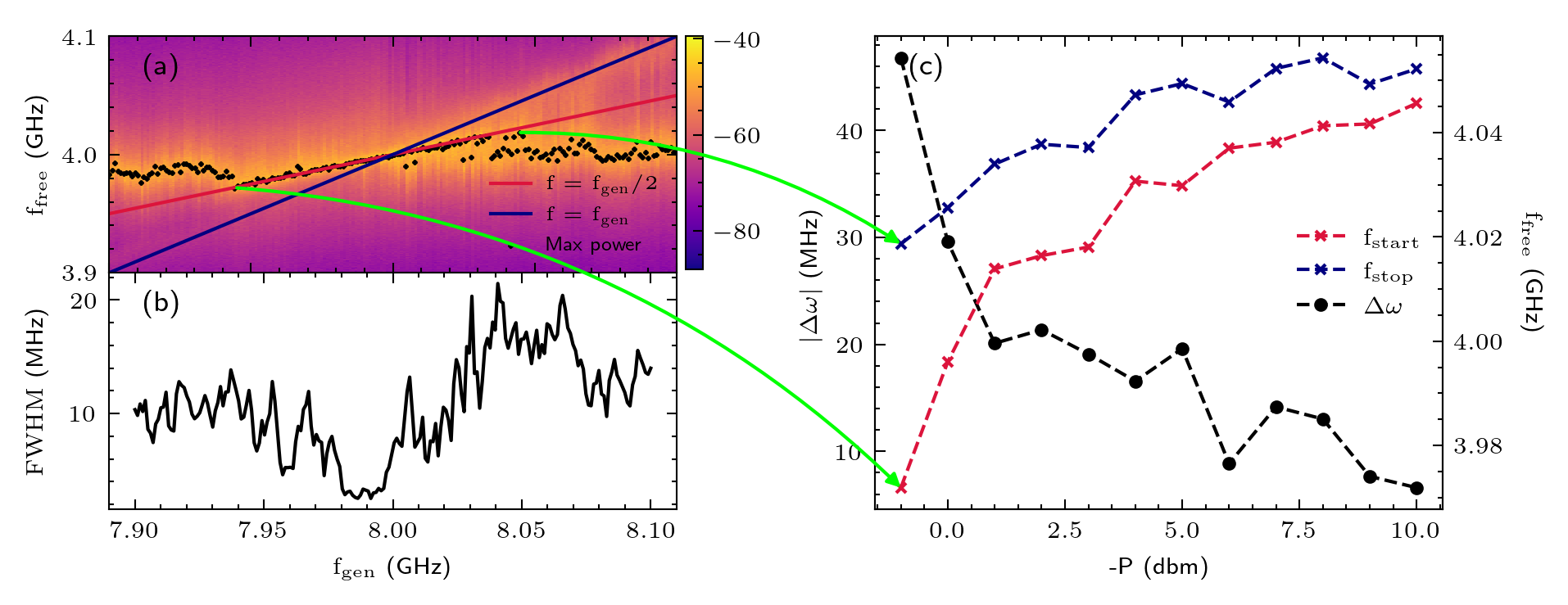}
    \caption{Experimental picture of an MTJ synchronizing to an external generator signal. (a) Power vs. frequency for different generator frequencies $f_{gen}$. Black dots represent maximum power points, the red line denotes $f_\textrm{gen}/2$, and the dark blue line $f_\textrm{gen}$. (b) FWHM of the signal generated by the STO versus $f_\mathrm{gen}$. The synchronization region is accompanied by a decrease in FWHM by a factor of 5. c) locking range $\Delta\omega$ as a function of generator power $P$. Note that the x-axis is in $-P$ units. As the generator power increases, the locking range increases too.}
    \label{fig:injection-spectra}
\end{figure*}

The connected element was supplied with a voltage of \SI{-250}{\milli\volt} and subjected to a magnetic field of \SI[per-mode = symbol, round-mode = places, round-precision=2]{-120}{\kilo\ampere\per\metre} at spherical angles of $\theta$ = \SI{25}{\degree} relative to the normal surface vector and in-plane $\phi$ = \SI{110}{\degree} measured from the positive x axis. This resulted in a free oscillation frequency of around \SI{4.075}{\giga\hertz}. Then the signal from the RF generator $f_{gen}$ is changed in the frequency range from \SI{7.9}{\giga\hertz} to \SI{8.3}{\giga\hertz}. The center frequency of the STO oscillation is presented as red dots overlayed on the spectra in Fig.\ref{fig:injection-spectra} and for a range of $f_{gen}$ values it follows the $f_{gen}/2$ frequency, indicating that the element is synchronized with the provided external $2f$ RF signal.

We calculate the FWHM for each measured point, shown in Fig.\ref{fig:injection-spectra}b). In the locking region, the FWHM significantly drops from the average of \SI{12.5}{\mega\hertz} to less than \SI{2.5}{\mega\hertz}. This drop in the FWHM of the signal is another indication of the synchronization process. 
We found that for a synchronization to occur, the experimental conditions must be very precisely controlled: the sample must be aligned precisely with an external magnetic field; the value of this magnetic field as well as DC bias require careful experimental tuning with successive approximation to achieve stable free oscillations.

\subsection{Power levels in synchronization to an external $2f$ signal}
The next step was to identify the power of the external generator necessary to achieve the synchronization of the two signals. To do so, we gradually decrease the generator power and measure the associated decrease in the locking range, with the power level determined precisely as follows. 
Using a vector network analyzer, the transmittance on the path from the RF signal generator to the connection port (see Fig.\ref{fig:measurement-setup}) was determined to be \SI{-10.692}{\dB} at the representative frequency of \SI{4.00}{\giga\hertz}. A similar measurement determined that the transmittance on the path from the connection port to the spectrum analyzer is \SI{15.833}{\dB} at the same frequency. The final path of the signal from the connection port to the sample is realized using an RF probe with flexible RF wires. At this point, there is an impedance mismatch between the \SI{50}{\ohm} system and the undetermined impedance of the sample. Therefore, the VNA was calibrated with an open, short, and reference \SI{50}{\ohm} impedance (Picoprobe Calibration Substrate CS-8) connected to the RF probe, and then a measurement was taken with a representative sample connected to the system. 
This allowed, after some standard and simple calculations, to apply a correction to all the measured power levels.
Synchronization, indicated by a decrease in FWHM, was determined at the injected power level delivered to the element of approximately \SI{-18}{\dBm}, while the true power generated by the element peaks at \SI{-29}{\dBm}. As a result, we conclude that the ability to lock the STO frequency to the external signal increases with increasing power of the external signal.
\end{appendix}

\end{document}